\documentclass[ reprint, amsmath, amssymb, aps, pra ]{revtex4-2}

\usepackage{graphicx}
\usepackage{bm}
\usepackage{amsmath,amssymb}
\usepackage{color}
\usepackage[caption=false]{subfig}
\usepackage{enumitem}

\usepackage{dcolumn}
\usepackage{hyperref}
\usepackage[mathlines]{lineno}

\usepackage{gensymb} 
\usepackage{makecell} 
\usepackage[normalem]{ulem} 
\usepackage{float}

\hypersetup{
    colorlinks,
    citecolor=black,
    filecolor=black,
    linkcolor=black,
    urlcolor=black
}

\begin{document}

\title{Bright, low-noise source of single photons at 780 nm  with improved phase-matching in rubidium vapor}

\author{Omri Davidson}
\affiliation{Department of Physics of Complex Systems, Weizmann Institute of Science, Rehovot 7610001, Israel}
\author{Ohad Yogev}
\affiliation{Department of Physics of Complex Systems, Weizmann Institute of Science, Rehovot 7610001, Israel}
\author{Eilon Poem}
\affiliation{Department of Physics of Complex Systems, Weizmann Institute of Science, Rehovot 7610001, Israel}
\author{Ofer Firstenberg}
\affiliation{Department of Physics of Complex Systems, Weizmann Institute of Science, Rehovot 7610001, Israel}

\begin{abstract}
\noindent
Future optical quantum networks could benefit from single photons that couple well to atoms, for realizing, \textit{e.g.}, quantum memories and deterministic photonic gates. However, the efficient generation of such photons remains a difficult challenge.
Recently, we demonstrated a bright multiplexed source of indistinguishable single photons with tunable GHz-bandwidth based on four-wave-mixing in rubidium vapor [Davidson \textit{et al.} 2021 New J. Phys. 23 073050]. Here we report on an improved implementation of this photon source. The new implementation employs a frequency-detuning regime that is better phase matched, a spatial-alignment procedure using single-mode fibers, a different rubidium isotope, and higher vapor-cell transmission. Characterization of the source is performed using superconducting-nanowire detectors with higher detection efficiency and lower jitter. Our source produces single photons with detected heralding efficiency of over 20\%, Hong-Ou-Mandel interference visibility of 88\%, generation rate of over 100 kilo-counts per second, and signal-to-noise ratio greater than 100, making it suitable for quantum information processing with photons.
\end{abstract}

\maketitle

\section{Introduction}
Single photons that can efficiently interact with alkali atoms are highly desirable for quantum optics experiments and for quantum information processing \cite{Eisman_2011_review_single_photons}. 
The performance of single-photon sources improved considerably in recent years, mainly by using spontaneous parametric down conversion (SPDC) \cite{2016_Pan_SPDC_10_photon_entanglement, 2018_Pan_SPDC_12_photon_entanglement,SPDC_source_2020_Takateo} and quantum dots \cite{photon_source_QD_Senellart_2016, 2016_Pan_QD, 2021_Tomm_QD_cavity}. However, efficiently generating and interfacing photons from these sources with atomic ensembles remains a difficult challenge \cite{2019_cavity_SPDC_review, 2020_quantum_storage_Chen, QD_atoms_interface_Akopian_2011, 2018_QD_interfacing_atomic_vapor_Michler}. 
Alternatively, single photons can be generated by atomic ensembles, making them inherently compatible with atomic systems \cite{2019_Du_cold_memory,2020_Laurat_memory_for_single_photons_cold_atoms}. Nevertheless, these sources typically operate at a low photon generation rate and moderate signal-to-noise ratio (SNR). 

Adopting the scheme developed by Lee \textit{et al.} \cite{photon_source_CW_hot_SebMoon_2016,2017_HOM_interference_two_sources_Seb_Moon,2018_source_CW_hot_atoms_Franson_interference_Seb_Moon}, we have recently demonstrated a multiplexed, heralded, single-photon source based on four-wave-mixing (FWM) in rubidium vapor \cite{Photon_source_paper}. The source employs the nearly Doppler-free ladder scheme $|5S_{1/2}\rangle - |5P_{3/2}\rangle - |5D_{5/2}\rangle $ of rubidium. It generates single photons at 780 nm, heralded by the detection of idler photons at 776 nm, with high rate and SNR. Multiplexing is achieved by operating two independent spatial channels producing indistinguishable photons, as verified by a Hong-Ou-Mandel (HOM) interference measurement. Additionally, by changing the optical depth (OD) of the atomic ensemble, the temporal width of the photons is tunable over a dynamic range of five. 

In this letter, we report on an improved experimental implementation of the photon source, based on the same principle setup. 
The improved source employs a frequency of the excitation fields which is better phase-matched, spatial alignment of the counter-propagating excitation beams at the single-mode level, a different rubidium isotope, and a vapor cell with higher transmission. We characterize the source using superconducting nanowire single-photon detectors (SNSPDs) with a higher detection efficiency and a lower detection time-jitter with respect to the detectors used in Ref.~\cite{Photon_source_paper}, which further improves the results.

\section{Experiment}
\begin{figure} 
	\centering
	\includegraphics[width=\columnwidth,trim=0.0cm 0.0cm 16.0cm 0.0cm ,clip=true] {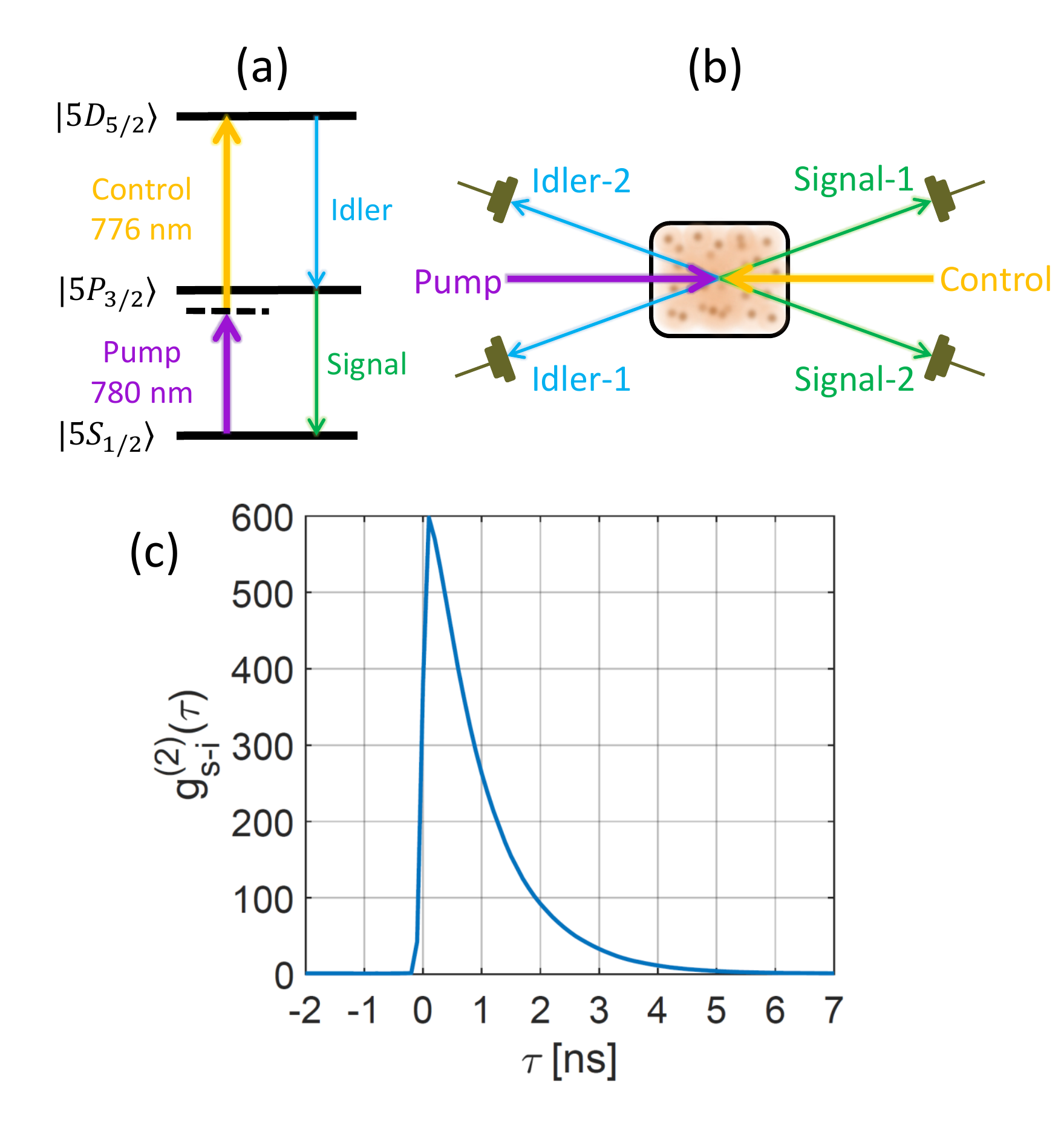}
	\caption{\textbf{Photon source experiment.} 
	(\textbf{a}) Atomic levels of rubidium used to generate the signal and idler photons in a FWM process.
    (\textbf{b}) The pump and control beams counter-propagate through the vapor cell to minimize the residual Doppler broadening of the two-photon transition. The signal and idler photons are emitted in the phase-matched directions. We spatially multiplex the photon source by collecting the signal and idler photons from both sides of the optical axis. 
    (\textbf{c}) Normalized cross-correlation $g^{(2)}_{\text{s-i}}(\tau)$ of the signal and idler photons versus their time separation $\tau$. 
    }
	\label{fig:experiment} 
\end{figure}

We start by briefly outlining the experimental scheme and refer the reader to Ref.~\cite{Photon_source_paper} for additional details. 
As shown in Fig.~\ref{fig:experiment}(a), the excitation fields, denoted as pump and control, couple the $|5S_{1/2},F=2\rangle\rightarrow |5P_{3/2},F=3\rangle$ and $|5P_{3/2},F=3\rangle\rightarrow |5D_{5/2},F=4\rangle$ transitions in $^{87}\text{Rb}$, respectively, with  transition wavelengths of 780 nm and 776 nm.
The combined two-photon transition is, on the one hand, nearly Doppler-free and, on the other hand, enables separation of the two wavelengths using standard thin-film interference filters. The signal and idler photons are generated in a FWM process from the respective transitions of the pump and control fields.

Figure~\ref{fig:experiment}(b) shows the geometry of the beams in the experiment. The pump and control beams are counter-propagating through a 25-mm-long vapor cell with isotopically purified $^{87}\text{Rb}$.
The signal and idler photons are emitted in the phase-matched directions and collected at an angle of $1.4\degree$ from the optical axis. As the phase-matching function has cylindrical symmetry, we collect the signal and idler photons from two sides of the optical axis and thus spatially multiplex our source, effectively creating two sources using the same vapor cell. 

The normalized bi-photon cross-correlation function, defined as $g^{(2)}_{\text{s-i}}(\tau) = \langle a_\text{i}^\dagger (t) a_\text{s}^\dagger(t+\tau) a_\text{s} (t+\tau) a_\text{i}(t)\rangle/[\langle a^\dagger_\text{s} a_\text{s}\rangle \langle a^\dagger_\text{i} a_\text{i} \rangle ]$, is shown in Fig.~\ref{fig:experiment}(c). Here $\langle \cdot \rangle$ denotes averaging over time $t$, and  $\tau$ is the time separation between signal and idler detections. A strong signal-idler correlation is evident, and the cascaded nature of the emission is seen in the asymmetric shape of the bi-photon cross-correlation. The background value of $g^{(2)}_{\text{s-i}}(\tau)$, due to uncorrelated signal and idler photons detection, is equal to 1. Therefore, its peak value $[g^{(2)}_{\text{s-i}}]_\text{max}$ is associated with the SNR of the source.

\section{Photon source improvements}
We now detail the changes made in the new implementation. First, we employ an isotopically-purified $^{87}\text{Rb}$ cell instead of the purified $^{85}\text{Rb}$ cell used previously. Except for the isotope, it is the same cell type and dimensions (same vendor, Precision Glassblowing). Before placing the cell in the setup, we immerse it in Acetone as a cleaning step and obtain a transmission of $96.5 \pm 1 \%$ at 780 nm in the optical setup (reported errors are 1 standard deviation).
The cell is heated to $\sim 45~\degree$C, such that $\text{OD}=4\pm 0.1$ [measured on the $|5S_{1/2},F=2\rangle \rightarrow |5P_{3/2},F=1,2,3\rangle$ transitions]. This OD is chosen to maximize the heralding efficiency of the source and to generate temporally-long photons while maintaining a good SNR \cite{Photon_source_paper}. 

The $^{87}\text{Rb}$ isotope has a larger hyperfine level splitting than $^{85}\text{Rb}$. Therefore, for a given atomic density, which determines the strength of the FWM interaction, the generated broadband signal photons are less absorbed in the $|5P_{3/2},F=1,2\rangle$ states ($|F=2,3\rangle$ in $^{85}\text{Rb}$). Using the numerical model described in Ref.~\cite{Photon_source_paper}, we estimate that the reduced absorption in $^{87}\text{Rb}$ increases the heralding efficiency and SNR for a given generation rate by $\sim 15\%$ compared to $^{85}\text{Rb}$ for our experimental parameters.

Second, we optimize the spatial alignment of the counter-propagating pump and control beams. In Ref.~\cite{Photon_source_paper}, the two beams were aligned by optimizing the transmission through two irises with a diameter of 0.8 mm, comparable to the beams' waist diameter of 0.9 mm. In the new implementation, we couple the outgoing control beam into the single-mode fiber of the incoming pump beam, thus verifying the mutual alignment of these beams at the single-mode level. This procedure improves the spatial mode-overlap of the pump and control beams by $15 \pm 2\%$ compared to that achieved with the irises, measured by the increase of the control power coupled into the incoming pump-beam fiber.

\begin{figure} 
	\centering
	\includegraphics[width=\columnwidth,trim=0.0cm 0.0cm 0.0cm 0.0cm ,clip=true] {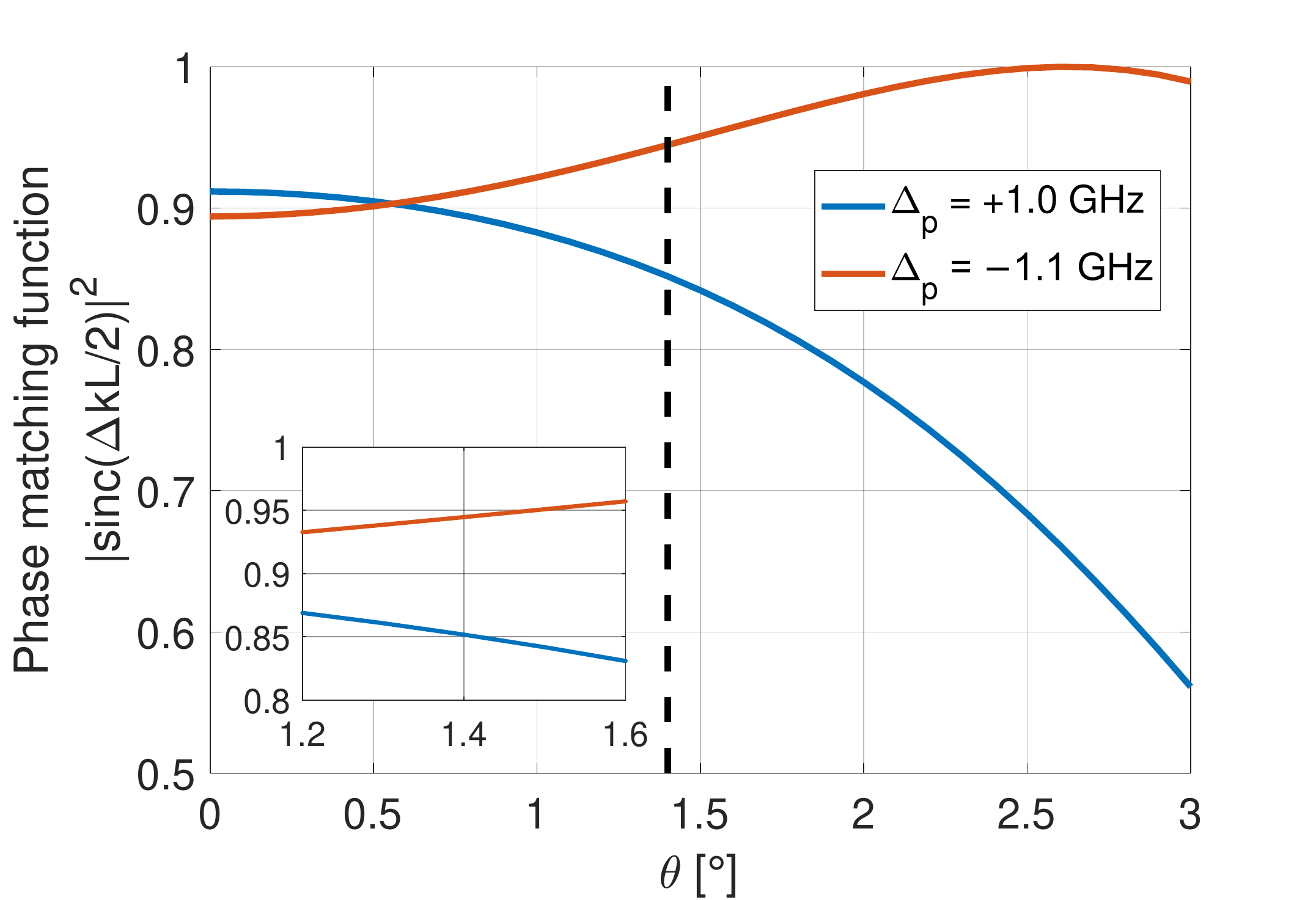}
	\caption{\textbf{Phase matching of the photon source.} 
    The phase-matching function $|\text{sinc}(\Delta k L/2)|^2$ for $L=25$ mm versus the angle of the signal beam $\theta$ from the optical axis, with the idler at the optimal angle of approximately $-\theta$. Positive pump detuning $\Delta_\text{p}=+1.0$ GHz (as in Ref.~\cite{Photon_source_paper}) is shown in blue, and negative pump detuning $\Delta_\text{p}=-1.1$ GHz (this work) in orange. The dashed black line indicates the collection angle in our implementation $\theta=1.4\degree$ (magnified in the inset). 
    }
	\label{fig:phase matching} 
\end{figure}

Third, we optimize the phase-matching of the FWM process. The bi-photon wavefunction amplitude depends on the phase-matching term $\psi(\tau) \propto \text{sinc}(\Delta kL/2)$ \cite{Du_2008_bi_photon_state_vector_model}, where $L$ is the length of the vapor cell and $\Delta k = (k_\text{p}-k_\text{c})-(k_\text{s}-k_\text{i})$ is the wavevectors mismatch. Here $k_\text{p}$, $k_\text{c}$, $k_\text{s}$, and $k_\text{i}$ are the projections (in absolute value) of the wavevectors of the pump, control, signal, and idler fields, respectively, on the optical axis. 
It follows that the bi-photon generation rate and the signal photon heralding efficiency scale as $|\text{sinc}(\Delta kL/2)|^2$. Note that this expression neglects the absorption (scattering) of the signal and idler photons, which alters the complex linear susceptibility and could be included in the model via the signal and idler wavevectors \cite{Du_2008_bi_photon_state_vector_model}.

Figure~\ref{fig:phase matching} shows the phase-matching function $|\text{sinc}(\Delta kL/2)|^2$ versus the angular deviation of the signal mode from the optical axis $\theta$ for a pump detuning of $\Delta_\text{p}=+1.0$ GHz (as in Ref.~\cite{Photon_source_paper}) and $\Delta_\text{p}=-1.1$ GHz used in this work. Here it is assumed that the signal and idler photons are emitted on resonance due to the third order susceptibility enhancement, as verified numerically using the model described in Ref.~\cite{Photon_source_paper}. We note that the angular deviation of the idler mode $\theta_\text{i}=\theta_\text{s}\times k_\text{s}/k_\text{i}$ differs from that of the signal by only $\sim 0.5\%$, due to momentum conservation in the transverse axis and the wavelength mismatch of the signal and idler photons.
As shown in Fig.~\ref{fig:phase matching}, it is possible to obtain perfect phase-matching $|\text{sinc}(\Delta kL/2)|^2=1$ with $\Delta_\text{p}<0$, but not with $\Delta_\text{p}>0$. This makes the previous choice $\Delta_\text{p}>0$ by us and others \cite{Photon_source_paper,photon_source_CW_hot_SebMoon_2016,2017_HOM_interference_two_sources_Seb_Moon,2018_source_CW_hot_atoms_Franson_interference_Seb_Moon} less preferable. We note however that near the phase-matching optimum, the difference between $\Delta_\text{p}<0$ and $\Delta_\text{p}>0$ increases quadratically with the vapor cell length, making the difference less significant in smaller cells \cite{photon_source_CW_hot_SebMoon_2016, 2017_HOM_interference_two_sources_Seb_Moon, 2018_source_CW_hot_atoms_Franson_interference_Seb_Moon}.
At the collection angle of $1.4\degree$ of our setup, previously designed to operate with $\Delta_\text{p}>0$, the phase-matching function with $\Delta_\text{p}=-1.1$ GHz is $11\%$ higher than with $\Delta_\text{p}=+1.0$ GHz. 

To understand why perfect phase-matching $\Delta k=0$ is achievable only with $\Delta_\text{p} <0$, we rewrite it as 
\begin{multline}
    \Delta k = \frac{1}{c}\Big[ (\omega_\text{SP}+\Delta_\text{p}) - (\omega_\text{PD}- \Delta_\text{p}) - \\ \big( \omega_\text{SP}\cos\theta_\text{s} - \omega_\text{PD}\cos(\theta_\text{s}\omega_\text{SP}/\omega_\text{PD}) \big) \Big],
\end{multline}
where $\omega_\text{SP}$ and $\omega_\text{PD}$ are the resonance frequencies of the $|5S_{1/2}\rangle \rightarrow |5P_{3/2}\rangle$ and the $|5P_{3/2}\rangle \rightarrow |5D_{5/2}\rangle$ transitions, respectively.
Expanding for small angles, we find 
\begin{equation}
    \Delta k \approx \frac{1}{c}\Big[ 2\Delta_\text{p} + \frac{1}{2}\theta^2_\text{s}\omega_\text{SP}(1-\omega_\text{SP}/\omega_\text{PD})  \Big].
\end{equation}
The second term in the brackets is positive for $\omega_\text{SP}<\omega_\text{PD}$, and therefore perfect phase-matching is only possible with $\Delta_\text{p}<0$.

To verify the improvement of phase-matching, we measure the bi-photon generation rate and heralding efficiency at $\Delta_\text{p}=\pm1.1$ GHz, in which the phase-matching function is $14\%$ higher in the negative detuning than in the positive detuning. This calculated factor agrees well with the increase we measure of $17 \pm 1\%$ in the bi-photon generation rate and $16 \pm 1\%$ in the heralding efficiency.

\begin{figure} 
	\centering
	\includegraphics[width=\columnwidth,trim=0.0cm 0.0cm 0.0cm 0.0cm ,clip=true] {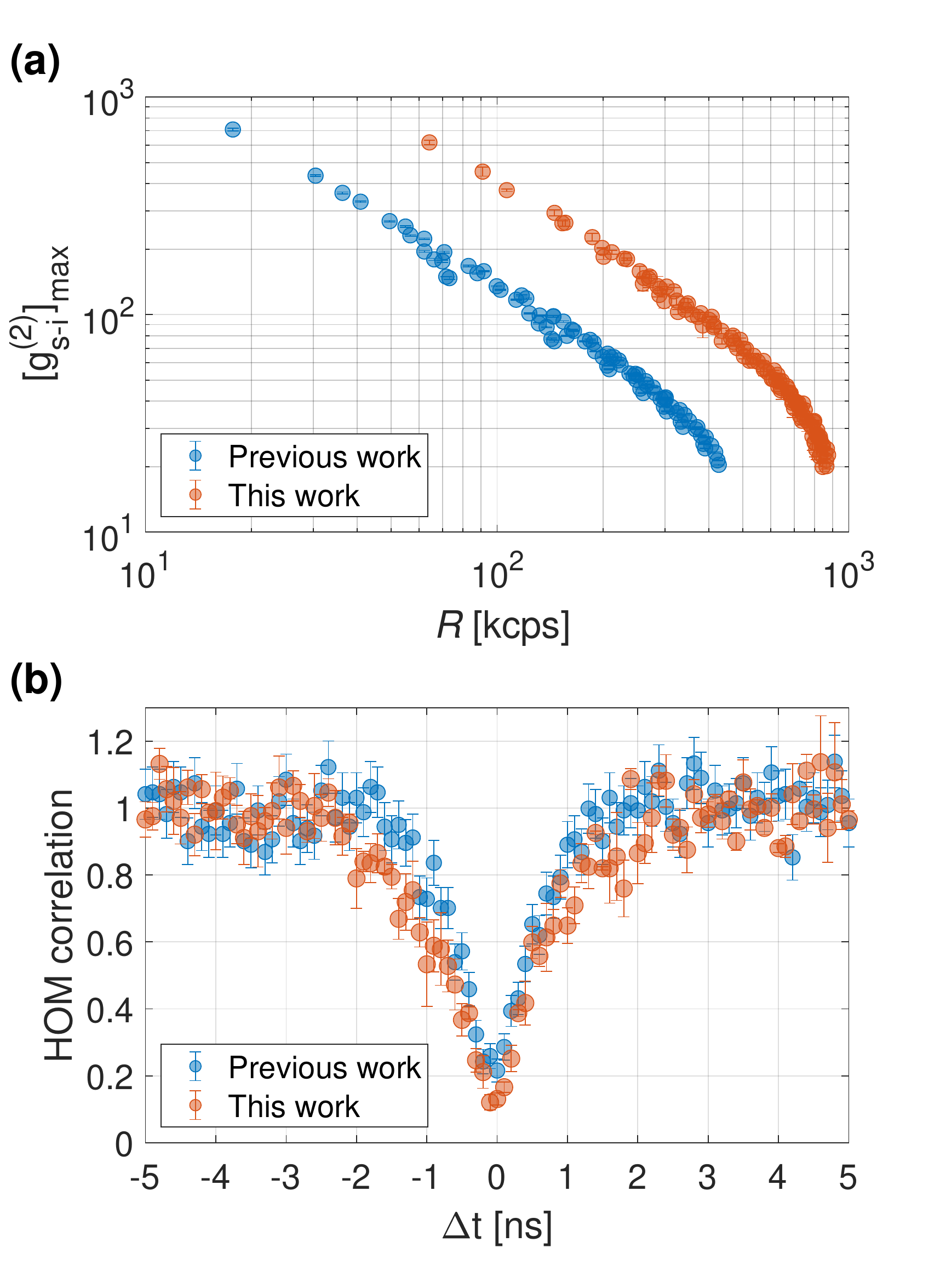}
	\caption{\textbf{Comparison of SNR and HOM correlation of the photon sources.} 
    (\textbf{a}) The peak of the bi-photon normalized cross-correlation $[g^{(2)}_{\text{s-i}}]_\text{max}$ versus the detected bi-photon generation rate $R$. The comparison is made for $\text{OD}=4$ in this work and $\text{OD}=9.3$ in Ref.~\cite{Photon_source_paper}.
    (\textbf{b}) Coincidence counts of the heralded signal-1 and signal-2 photons after interference on a beam splitter versus the time difference between the detection of the heralding idler photons. We consider a heralding event if the signal and idler photons were detected within a 3.5 ns-long detection time window, which accounts for more than 95\% of the heralded photons' energy.
    In (\textbf{a}) and (\textbf{b}), results from the new implementation (orange) are compared to those from Ref.~\cite{Photon_source_paper} (blue).
    }
	\label{fig:comparsion} 
\end{figure}

Lastly, we characterize the photon source using SNSPDs with a detection efficiency of $\sim 90\%$ and detection time jitter full-width at half-maximum of $55$ ps, compared to $\sim 68\%$ and $\sim 350$ ps of the detectors used in Ref.~\cite{Photon_source_paper}. 
Low detection-time jitter prevents smearing of the peak of
$g^{(2)}_{\text{s-i}}(\tau)$, thus increasing the SNR of the source, and also increases the heralded photons’ purity \cite{2015_Du_photon_purity}, thus increasing the HOM interference visibility \cite{2008_Mosley_JSA_and_purity}.

Higher detection efficiency of the detectors improves the measured heralding efficiency $\eta_\text{h}$ and the detected bi-photon generation rate $R$. While $R$ can be increased by stronger pumping, the SNR for a given $R$ is nonetheless improved. This can be understood as follows: 
In bi-photon sources based on parametric processes, such as SPDC and FWM, the SNR is decreased as the rate is increased \cite{cavity_SPDC_Treutlein_2020, Photon_source_paper, 2022_Seb_Moon_cesium_photon_source} due to an increase in the multi-pair emission probability. Even if the detectors have a finite detection efficiency $0<\eta_\text{d}<1$, the SNR is the same as that with perfect detectors, as evident from the definition of $g^{(2)}_{\text{s-i}}(\tau)$. However, as $R \propto \eta_\text{d}^2$, higher detection efficiency improves the SNR for a given generation rate. It follows, as intuitively expected, that high-performance photon sources require high detection efficiency.

\section{Comparison of performance}
Figure~\ref{fig:comparsion}(a) shows the SNR ($[g^{(2)}_{\text{s-i}}]_\text{max}$) versus $R$ in kilo-counts per second (kcps). It is evident that the SNR is significantly improved in this work. 
As an example, for a bi-photon generation rate of $R=200$ kcps, we obtain a three-fold increase from $[g^{(2)}_{\text{s-i}}]_\text{max}=64\pm 1$ in Ref.~\cite{Photon_source_paper} to $[g^{(2)}_{\text{s-i}}]_\text{max}=202\pm 1$ in this work.
Increasing the OD will further increase the SNR, at the cost of lowering the heralding efficiency \cite{Photon_source_paper}. 
For completeness, we note that as in Ref.~\cite{Photon_source_paper}, the SNR for a given generation rate of the second spatial channel is slightly lower than that of the first channel (presented here) mainly due to the increased scattering noise from the control field.

We show the heralded signal-1 and signal-2 coincidence counts in a HOM interference setup \cite{HOM_1987} in Fig.~\ref{fig:comparsion}(b). Here, we consider a heralding event of the signal photons if the signal and idler photons are detected within a 3.5 ns-long detection time window. This time window captures over 95\% of the heralded photons' energy. The raw HOM visibility is increased from $V=78\pm 3 \%$ in Ref.~\cite{Photon_source_paper} to $V = 88\pm 2 \%$ in this work with similar OD and SNR. We note that here the reported visibility of the previous photon source is lower than the value reported in Ref.~\cite{Photon_source_paper} due to the larger coincidence detection time window considered here as a coincident signal-idler photons detection.
  
A summary of the sources performance is shown in Table \ref{tab:photon source comparison}. 
It compares the sources SNR and HOM visibility discussed above, as well as the sources heralding efficiency. The measured heralding efficiency improves from $\eta_\text{h}=10.5 \pm 0.1\%$ in Ref.~\cite{Photon_source_paper} to $\eta_\text{h}=24 \pm 0.1\%$ in this work.

To quantify the different contributions of the photon source improvements, we consider the heralding efficiency. The improved detectors increase $\eta_\text{h}$ by $\sim 40\%$ (including the fiber-to-air interface), and the improved phase-matching increases $\eta_\text{h}$ by $\sim 11\%$. The remaining $\sim 50\%$ improvement originates from the pump beam alignment, higher vapor cell transmission, and different rubidium isotope.

\begin{table}[t]
\caption{\textbf{Comparison of this work and Ref.~\cite{Photon_source_paper}.} }
\label{tab:photon source comparison}
\begin{tabular}{|l|c|c|c|}
\hline
 & \makecell{Heralding \\ efficiency \\ $\eta_\text{h}$} 
 & \makecell{HOM \\ visibility  \\ $V$} 
 & \makecell{ Max normalized  \\ cross-correlation \\ $[g^{(2)}_\text{s-i}]_\text{max}$ \\ @ $R=200$ kcps }  
 \\ \hline
\makecell{This \\  work  }    & $24\% \pm 0.1\%$                               & $88\% \pm 2\%$                   & $202\pm 1$                                                                                 \\ \hline
\makecell{Ref.~\cite{Photon_source_paper} } & $10.5\% \pm 0.1\%$                              & $78\% \pm 3\%$                   & $64\pm 1$                                                                                  \\ \hline
\end{tabular}
\end{table}

We verify the single-photon nature of the generated photons by measuring the auto-correlation of the signal photons $g_\text{c}^{2}(0)$, conditioned on the detection of an idler photon. Using a 3.5~ns-long detection time window we measure $g_\text{c}^{2}(0)=0.0112\pm 0.0001$ for $R=37$~kcps.
As a comparison, in Ref.~\cite{Photon_source_paper}, we obtained a similar value $g_\text{c}^{2}(0)=0.012\pm 0.0003$ for $R=15$~kcps, with $\text{OD}=9.3$ and a 2.5~ns-long detection time window that accounts for 95\% of the photons' energy with the higher OD.

\section{Discussion}
The advantage of heralded single-photons that are generated using an atomic ensemble is that they can be stored in an atomic quantum memory based on the same ladder-level scheme \cite{FLAME_paper, FLAME_2_paper}. Indeed, our upgraded photon source, combined with a quantum memory \cite{FLAME_2_paper}, has been used to construct synchronized two-photon states \cite{Photon_synchronization_paper}. Notably, high-rate and low-noise photon generation with high heralding efficiency is required to efficiently synchronize the photons for multi-photon state generation \cite{Photon_synchronization_paper}. 

In conclusion, we demonstrate a heralded single-photon source with improved performances compared to our original implementation~\cite{Photon_source_paper}. Our photon source generates indistinguishable heralded single photons with high rate and SNR. It can readily be used as a source of single photons for systems based on rubidium atoms, making it a valuable resource for a wide range of applications in the field of quantum optics, including those requiring multi-photon states.

\begin{acknowledgments}
We acknowledge financial support from the Israel Science Foundation, the US-Israel Binational Science Foundation (BSF) and US National Science Foundation (NSF), the Minerva Foundation with funding from the Federal German Ministry for Education and Research, the Estate of Louise Yasgour, and the Laboratory in Memory of Leon and Blacky Broder.
\end{acknowledgments}


\bibliographystyle{unsrt}

\bibliography{Photon_source_2_bibliography}

\end{document}